# Highly dispersed Ru nanoparticles anchored on NiAl layered double oxides catalyst for selective hydrodeoxygenation of vanillin

Yongjian Zeng[a], Lu Lin[b], Di Hu[a], Zhiwei Jiang[a], Shaimaa Saeed[d], Ruichao Guo[a], Ibrahim Ashour[e], Kai Yan[a, c] *

[a] Guangdong Provincial Key Laboratory of Environmental Pollution and Remediation Technology, School of Environmental Science and Engineering, Sun Yat-sen University, 135 Xingang Xi Road, Guangzhou, 510275, China

[b] School of Materials Science and Engineering, State Key Laboratory of Optoelectronic Materials and Technologies, Key Lab of Polymer Composite & Functional Materials, Sun Yat-sen University, 135 Xingang Xi Road, Guangzhou, 510275, China

[c] Guangdong Laboratory for Lingnan Modern Agriculture, South China Agricultural University, Guangzhou, 510642, China

[d] Chemical Engineering Department, Tanta Higher Institute of Engineering and Technology, Tanta, Egypt

[e] Chemical Engineering Department, Faculty of Engineering, Minia University, Tanta, Egypt

*Corresponding Author Email: yank9@mail.sysu.edu.cn

**Abstract:** The hydrodeoxygenation (HDO) of lignin-derived feedstocks into value-added chemicals with high efficiency and selectivity is desirable for the utilization of biomass resource. The complex oxygen-containing groups of lignin-derived substance result in the challenge of the low selectivity toward the required product. In this work, highly dispersed Ru nanoparticles anchored on $Ni_3Al_1$ layered double oxides (LDOs) catalyst derived from NiAl layered double hydroxides (LDHs) with flower-shaped morphology was constructed by a simple deposition-reduction method. The introduction of LDHs-derived support can significantly impact the catalytic activity for the HDO of lignin-derived vanillin (VL) into 2-methoxy-4-methylphenol (MMP). The Ru/$Ni_3Al_1$-400 catalyst obtained complete conversion of VL and 94.2% yield of MMP at 130 °C in methanol solvent, much better than the catalysts without LDHs-derived support. The methanol solvent is beneficial for the conversion of reaction intermediate of vanillin alcohol (VA). Detailed characterization reveals that the existence of the enhanced metal-support interaction over Ru/$Ni_3Al_1$-400 and the easily accessible acid sites facilitate the production of MMP.

# 1. Introduction

The issues of increasing fossil energy shortage [1, 2] and global warming [3] have boosted the worldwide demand for sustainable and economical resource [4, 5] such as biomass, the only renewable carbon resource that can produce chemicals and fuels [6]. Lignin is one of the main constituents of lignocellulosic biomass, which could provide abundant source of aromatic compounds [7]. The efficient utilization of lignin is believed to be a promising strategy to address the current energy crisis. However, the complex structure and rich oxygen-containing groups of lignin hinder its valorization into value-added chemicals [8, 9]. Hydrodeoxygenation (HDO) of biomass-derived derivatives has been widely considered as an efficient and feasible strategy to remove excess oxygen content in lignin [10, 11].

2-methoxy-4-methylphenol (MMP) is a multifunctional intermediate for the production of pharmaceuticals, fragrances, and biofuel [12], which can be obtained from the lignin-derived bio-oil vanillin (VL) via HDO process [13]. The conversion of VL to MMP has been extensively investigated over heterogeneous metal catalysts. Small-sized metal particles are usually beneficial to achieve high yield of MMP. Santos et al reported that the HDO activity of Pd/C, Au/C and Ru/C increased with the increasing metal dispersion [14]. Lu and co-workers prepared atomically dispersed Pd single atoms and clusters over the SAPO-31 support via a photochemical method. The as-obtained Pd$_{SA+C}$/SAPO-31 afforded exceptionally higher turnover frequency (3000 h$^{-1}$) than Pd nanoparticles catalysts (200 h$^{-1}$) for the VL-to-MMP transformation [15]. In view of the dehydroxylation process during HDO reaction, the synergic effect between metal center and acid sites has been recognized to facilitate the polarization and cleavage of C-O bonds in VL molecules [16]. This motivates us to combine small-sized metal centers and the cooperated acid sites in one catalyst so as to improve the selectivity of MMP for the HDO of VL.

Layered double hydroxides (LDHs) with a two-dimensional structure [17] have been widely recognized as a type of excellent precursor for constructing heterogeneous catalysts due to their flexible and adjustable composition [18-22]. More importantly, the topological transformation of LDHs into layered double oxides (LDOs) is likely to induce strong interactions between metal and support, which could improve the dispersion of metal nanoparticles [23-25], expose more active sites, and accelerate the mass transfer of reactants [26]. Besides, acid sites could also be readily formed in the LHDs-derived LDOs supports. Recently, LDHs-based materials have attracted much attention in the catalytic biomass conversion. Feng et al. used $Ni_1MgAl$-LDHs derived $Ru/Ni_1MgAlO_x$ catalyst for the reductive amination of furfural and obtained 91.3% yield of furfurylamine [27]. Wang and co-workers reported that 0.4%$Pt/Co_2AlO_4$ catalyst derived from LDHs precursor could afford 99% yield of 2,5-dimethylfuran from 5-(hydroxymethyl) furfural via HDO process [28]. Therefore, LDHs are promising to be the suitable precursors to construct bifunctional catalysts with both small-sized metal centers and acid sites for the conversion of VL to MMP via HDO process.

Herein, an efficient $Ru/Ni_3Al_1$ catalyst was prepared via precipitation-impregnation of Ru on NiAl-LDHs and the following calcination and reduction process for the HDO of VL. Thorough characterizations were employed to evidence the exquisite structure of Ru/Ni$_3$Al$_1$, in which well dispersed Ru species were anchored on flower-like Ni$_3$Al$_1$-LDOs with easily accessible acid sites. Compared with the self-prepared, benchmark Ru/Al$_2$O$_3$ catalyst, the Ru/Ni$_3$Al$_1$ showed much higher yield of MMP (94.2%) from the HDO of VL. The reaction mechanism for the production of MMP over Ru/Ni$_3$Al$_1$ with methanol as solvent was deeply understood, and the stability was also investigated. This work presents an efficient catalyst synthesis approach for the fabrication of metal-acid bifunctional materials, and shows a prominent example of enhancing the HDO activity of catalysts for biomass conversion.

## 2. Experimental section

### 2.1. Catalyst preparation

#### 2.1.1 NiAl-LDHs precursor preparation

The NiAl-LDHs precursor was prepared by using a hydrothermal method. Typically, 22.5 mmol of Ni(NO$_3$)$_2$·6H$_2$O, and 7.5 mmol of Al(NO$_3$)$_3$·9H$_2$O with molar radio of 3:1 were dissolved in 50 mL of ultrapure water with 120 mmol of urea. The as-obtained homogenous solution was sealed into a Teflon-lined autoclave, and heated at 160 °C for 8 h. The autoclave was cooled to room temperature in flowing water after hydrothermal treatment. The solid was separated from the resulting suspension, and washed with ultrapure water until neutral by centrifugation at 10000 rpm. The LDHs precursor was obtained by drying at 80 °C for 8 h, and named as Ni$_3$Al$_1$-LDHs.

For comparison, Al(OH)$_3$ precursor was prepared according to the above steps except without the addition of Ni(NO$_3$)$_2$·6H$_2$O.

### 2.1.2 Ru-based catalysts preparation

Ru-based catalysts were synthesized via a precipitation-deposition method. In a typical procedure, the $Ni_3Al_1$-LDHs was dispersed into 100 mL of ultrapure water, and a certain amount of $RuCl_3$ solution (1 mg/mL) was added dropwise to the suspension. After stirring for 20 min, the pH of suspension was adjusted into 8 by a mixed solution containing 0.002 M $Na_2CO_3$ and 0.008 M NaOH with stirring. Then, the suspension continued to stir for 3 h. The solid was separated, and washed with ultrapure water by centrifugation at 10000 rpm for several times, followed by drying at 80 °C for 10 h. The powder was calcined in air at 600 °C for 2 h to obtain $RuO_x/Ni_3Al_1$, and then reduced in 10% v/v $H_2$/Ar mixed gas at 400 °C for 4 h to gain $Ru/Ni_3Al_1$-400.

In comparison, $Ru/Al_2O_3$-400 was prepared according to the above steps except the $Ni_3Al_1$-LDHs was replaced by $Al(OH)_3$ precursor. Besides, $Ni_3Al_1$-LDOs was prepared by the calcination of $Ni_3Al_1$-LDHs in air at 600 °C for 2 h, and then reduced in 10% v/v $H_2$/Ar mixed gas at 400 °C for 4 h to gain Ru-free catalyst, denoted as $Ni_3Al_1$-400.

### 2.2. Characterization

X-ray diffraction (XRD) patterns for analyzing the crystal structure of the catalysts were recorded on a Rigaku Ultima IV X-ray diffractometer equipped with Cu Kα radiation in the scanning range of 2 theta from 5° to 90° at 40 kV and 40 mA. Nitrogen adsorption/desorption experiments were performed on a Micromeritics ASAP 2020 for determining the specific surface areas of the catalysts via Brunauer-Emmett-Teller (BET) model. The metal contents of the catalysts were determined by an inductively coupled plasma optical emission spectrometer (ICP-OES). Microwave digestion was used for the mineralization of Ru. A certain amount of aqua regia and hydrofluoric acid were added to the container containing the material, and a pre-digestion (120 °C for 0.5 h) was carried out before the microwave digestion (Heating at 130 °C for 3 min, then heated to 150 °C for 10 min, and heated to 180 °C for 30 min). Scanning electron microscope (SEM) and transmission electron microscopy (TEM) were conducted on a Gemini500 and FEI Tecnai G2 F30, respectively, for characterizing the morphology of the catalysts. X-ray photoelectron spectroscopy (XPS) measurements were conducted on a Thermo Fisher Scientific ESCALAB 250 spectrometer using Al Kα radiation for analyzing the valence state of the elements in the catalysts. The C 1s peak located on 284.8 eV was used to calibrate. $H_2$ temperature-programmed reduction ($H_2$-TPR) and $NH_3$ temperature-programmed desorption ($NH_3$-TPD) experiments were performed on a Micromeritics AutoChem II 2920 instrument for determining the reducibility and acidity of the prepared catalysts, respectively. As for $H_2$-TPR test, 50 mg of the sample was dried in 50 mL/min of He flow for 1 h at 200 °C, and cooled to 50 °C. Then, the sample was heated in 30 mL/min of 10% v/v $H_2$/Ar flow from 50 °C to 800 °C with a heating rate of 10 °C/min. As for $NH_3$-TPD test, 50 mg of the sample was pretreated in He flow for 1 h at 200 °C, and cooled to 50 °C. Then, the sample was saturated with 30 mL/min of 10% v/v $NH_3$/He flow for 30 min followed by purging with He flow for 1 h. After that, the sample was heated in 30 mL/min of He flow from 50 °C to 800 °C with a heating rate of 10 °C/min.

## 2.3. Catalytic reaction

The HDO reaction of VL was carried out in a 25 mL of stainless-steel autoclave. Typically, 1 mmol of vanillin, 5 mL methanol (solvent), dodecane (internal standard) and 20 mg catalyst were added into the autoclave. The air in the autoclave was displaced by $H_2$ for 3 times. Then, the autoclave was charged with 2 MPa $H_2$, and heated at 130 °C for 8 h with stirring at 1000 rpm. The autoclave was cooled to room temperature in flowing water after reaction. The reaction solution was analyzed by gas chromatography (Techcomp 7900). The reaction products were identified by GC-MS (Agilent 7890B-5977B). Each catalytic reaction was tested three times to ensure the data repeatability.

The conversion of VL and the yield of products were calculated as the following:

$$\text{Conv. (\%)} = \frac{\text{moles of the added VL - moles of the residual VL}}{\text{moles of the added VL}} \times 100\%$$

$$\text{Yield (\%)} = \frac{\text{moles of a product}}{\text{moles of the added VL}} \times 100\%$$

## 3. Results and discussion

### 3.1. Synthesis and structural characterization.

The synthesis procedure used to fabricate Ru/Ni$_3$Al$_1$ catalyst is shown in Fig. 1a. The loading of Ru over the catalysts was about 1 wt.% measured by ICP-OES (Table S1). The Ru metal nanoparticles were uniformly dispersed on Ni$_3$Al$_1$-LDHs-based support after precipitation-deposition and following calcination and reduction process. Fig. 1b showed the XRD patterns of different samples. The characteristic diffraction peaks of Ni$_3$Al$_1$-LDHs were observed at 2$\theta$ of 11.6, 23.5, 35.1, 39.6, 47.2, and 61.2°, corresponding to the (003), (006), (012), (015), (018), and (110) crystal planes of LDHs phase (JCPDS No. 15-0087) [29]. It indicated that $Ni_3Al_1$-LDHs precursor was prepared successfully via the hydrothermal method. There were only NiO characteristic diffraction peaks (JCPDS No. 47-1049) at $2\theta$ of 37.4, 43.6, 63.4, 75.7, and 80.2° observed on the XRD pattern of $Ni_3Al_1$-400 prepared by the calcination and reduction of LDHs precursor. The XRD diffraction peaks of Ru/$Ni_3Al_1$-400 had no difference from that of $Ni_3Al_1$-400, indicating that the structure of the $Ni_3Al_1$-LDHs-based support had limited changes after the deposition of Ru nanoparticles. The absence of the diffraction peaks of metal Ru showed that the Ru nanoparticles were uniformly dispersed on the LDHs-based support. On the contrary, $Al_2O_3$ supported Ru catalyst (Ru/$Al_2O_3$-400) presented the diffraction peaks of metal Ru (JCPDS No. 06-0663) at $2\theta$ of 38.4, 42.2, 44.0, 58.3, and 78.4° [30] and $Al_2O_3$ support simultaneously (Fig. S1a). The difference between the XRD results of Ru/$Ni_3Al_1$-400 and Ru/$Al_2O_3$-400 revealed that the LDHs-based support was beneficial for the dispersion of Ru nanoparticles. Fig. 1c-e showed the SEM images of the LDHs precursor and the prepared catalysts. Both $Ni_3Al_1$-LDHs (Fig. 1c) and $Ni_3Al_1$-400 (Fig. 1d) possessed hierarchical nanosheets stacking flower-shaped morphology. The flower-shaped morphology was also retained after the addition of Ru metal phase to LDHs-based support (Fig. 1e). It is possible that the hierarchical nanosheets stacking flower-shaped morphology of LDHs-based support might increase the distribution of Ru nanoparticles and the accessibility between the reactant and the active sites [26]. The SEM image of Ru/$Al_2O_3$-400 is shown in Fig. S1b, which is composed of the irregular nanosheets. The $N_2$ adsorption-desorption isotherms of Ru/$Ni_3Al_1$-400 and Ru/$Al_2O_3$-400 are shown in Fig. S2. The hysteresis loop with H3 of type IV isotherm for the two catalysts indicated the plate-like structure [31]. The specific surface area of the two catalysts was similar (Table S1).

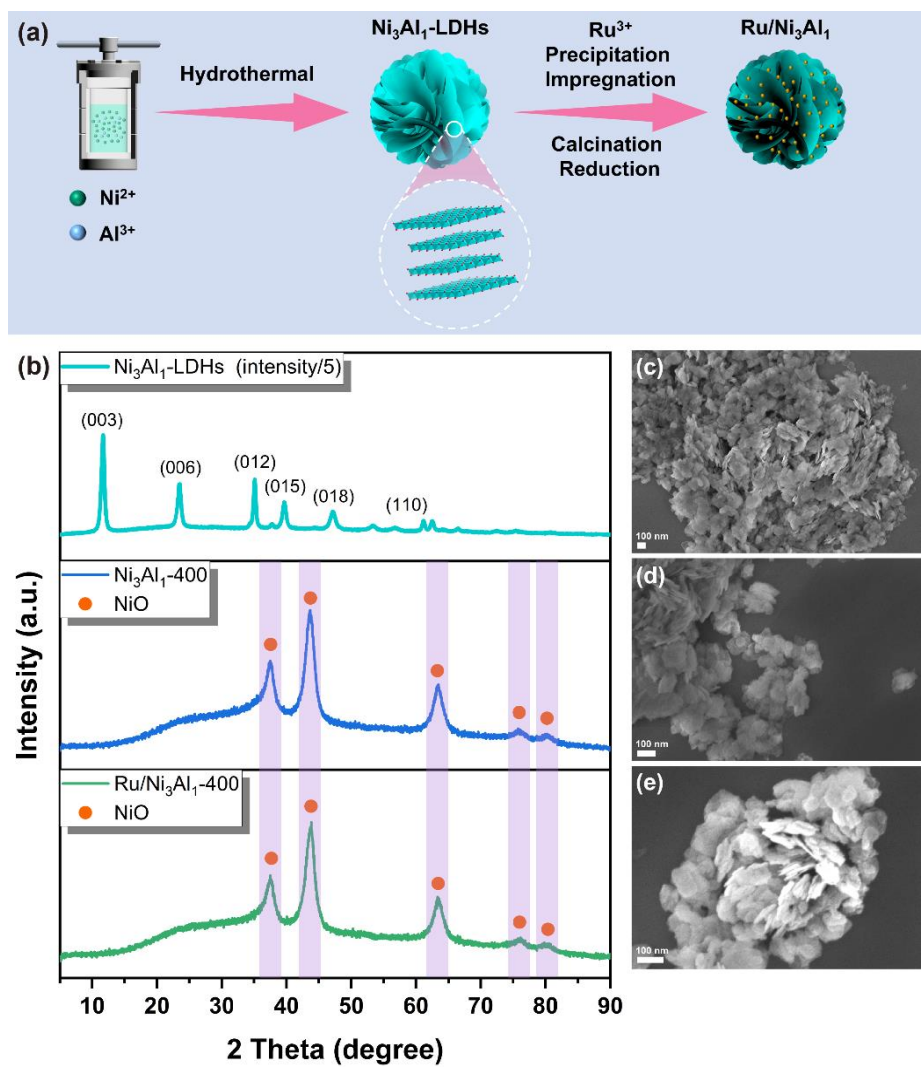

**Fig. 1.** Synthesis procedure of Ru/Ni$_3$Al$_1$ catalyst (a). XRD of Ni$_3$Al$_1$-LDHs, Ni$_3$Al$_1$-400, and Ru/Ni$_3$Al$_1$-400 catalysts (b). SEM images of Ni$_3$Al$_1$-LDHs (c), Ni$_3$Al$_1$-400 (d), and Ru/Ni$_3$Al$_1$-400 (e).

The morphology and structure of the prepared catalysts were further examined by high-resolution TEM analysis. Fig. 2a1 exhibits the structure of hierarchical nanosheets stacking for Ni$_3$Al$_1$-400. The crystal plane spacing of 0.209 nm corresponded to (200) plane of NiO (Fig. 2a2). Noteworthily, only the crystal plane spacing assigned to (200) and (111) planes of NiO are presented in the TEM image of Ru/Ni$_3$Al$_1$-400 and no visible Ru nanoparticles were detected (Fig. 2b2), indicating the highly dispersed Ru species in Ru/Ni$_3$Al$_1$-400. For Ru/Al$_2$O$_3$-400, the aggregation of Ru nanoparticles could be observed obviously in Fig. 2c1, and the crystal plane spacing of 0.122 nm corresponded to (103) plane of Ru is showed in Fig. 2c2. In order to further investigated the distribution of Ru among Ru/Ni$_3$Al$_1$-400, the element mapping was carried out. As shown in Fig. 2d, it clearly depicted the uniform distribution of Ru element on the surface of LDHs-based support. The results of TEM and element mapping were in agreement with the XRD results, confirming the key role of LDHs precursor for the introduction of Ru nanoparticle with high dispersity on the support.

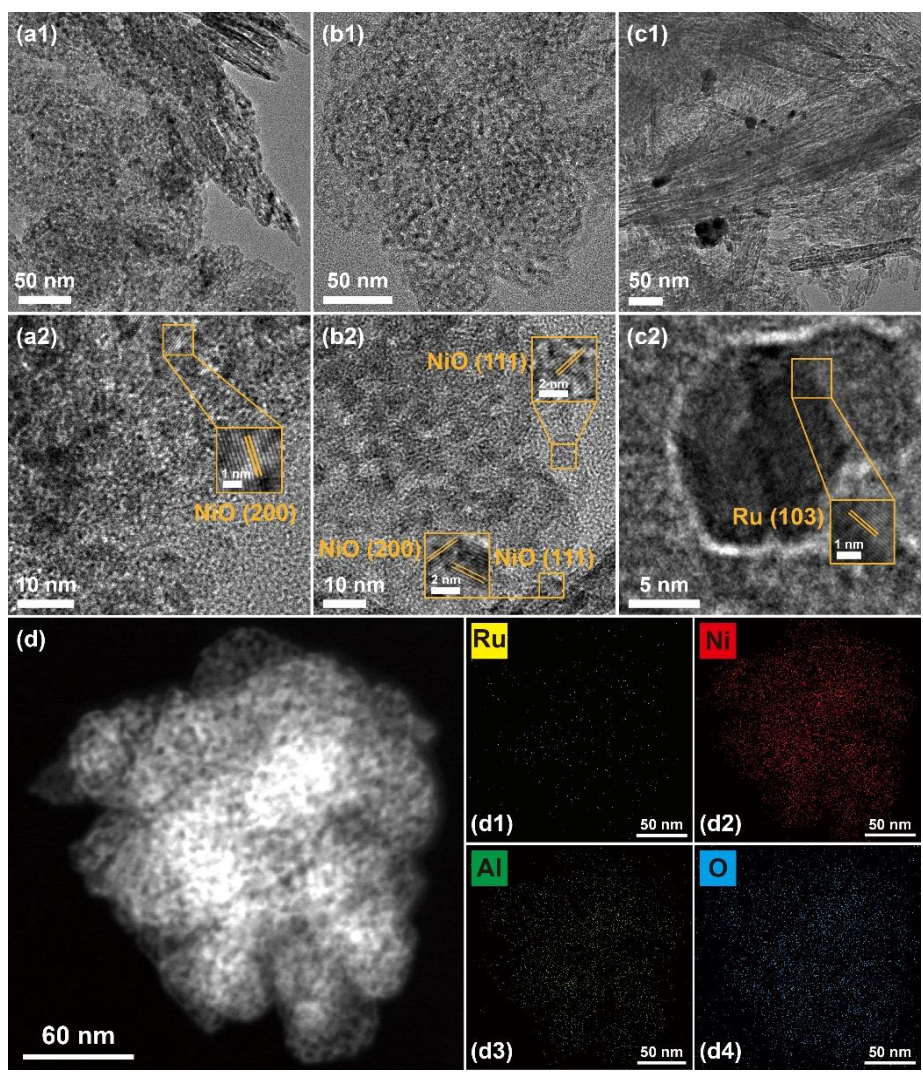

**Fig. 2.** TEM images of Ni$_3$Al$_1$-400 (a), Ru/Ni$_3$Al$_1$-400 (b), and Ru/Al$_2$O$_3$-400 (c). HAADF-STEM and element mapping images of Ru/Ni$_3$Al$_1$-400 (d).

The chemical composition and surface electronic structure of Ni$_3$Al$_1$-400, Ru/Ni$_3$Al$_1$-400, and Ru/Al$_2$O$_3$-400 were evaluated by XPS (Fig. 3). Fig. 3a shows the high-resolution XPS spectra of Ni 2p orbit for Ni$_3$Al$_1$-400 and Ru/Ni$_3$Al$_1$-400. For the two catalysts, there were two signals at around the binding energy of 854.5 and 860.7 eV after the deconvolution of Ni 2p$_{3/2}$ spectra, corresponding to the characteristics of Ni$^{2+}$ specie and its satellite peak [32]. It indicated that the form of nickel existed in Ni$_3$Al$_1$-400 and Ru/Ni$_3$Al$_1$-400 were nickel oxides, agreeing with the result of XRD.

The Ru 3p XPS profile of Ru/Al$_2$O$_3$-400 and Ru/Ni$_3$Al$_1$-400 are shown in Fig. 3b. For Ru/Al$_2$O$_3$-400, the fine Ru 3p$_{3/2}$ spectra presented two fitted peaks at about the binding energy of 460.8 and 462.9 eV, associating with the existence of Ru$^0$ and Ru$^{4+}$ species, respectively [33]. It was obvious that the fitted Ru$^0$ and Ru$^{4+}$ peaks of Ru/Ni$_3$Al$_1$-400 exhibited slight positive shifts compared with Ru/Al$_2$O$_3$-400, which located at 461.8 and 463.1 eV, respectively. The changes in the binding energy of Ru species for Ru/Ni$_3$Al$_1$-400 could be attributed to the decrease in the electron density of Ru atoms caused by the electron transfer from Ru to the LDHs-based support [34, 35], proving the existence of the enhanced interaction between metal and support [36], which was beneficial to improve the dispersion of Ru nanoparticles.

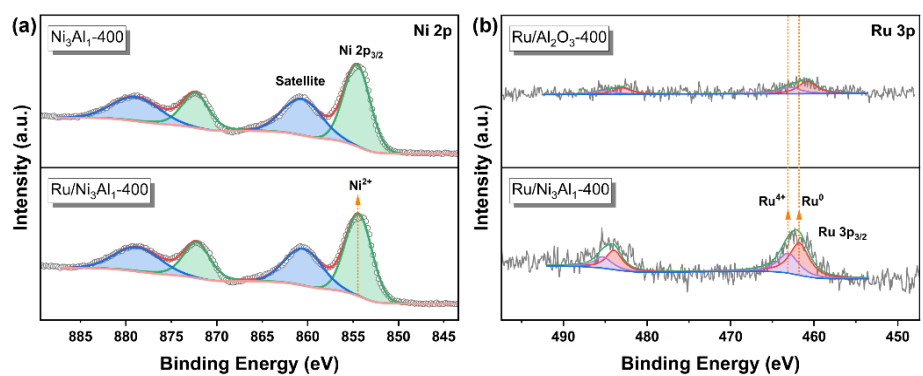

**Fig. 3.** XPS spectra of the Ni 2p for Ni$_3$Al$_1$-400 and Ru/Ni$_3$Al$_1$-400 (a). XPS spectra of the Ru 3p for Ru/Al$_2$O$_3$-400 and Ru/Ni$_3$Al$_1$-400 (b).

The reducibility of the calcinated samples was investigated by H$_2$-TPR test (Fig. 4a). The strong reduction peak of Ni$_3$Al$_1$-LDOs located at 656 °C was ascribed to the reduction of nickel oxide. After the introduction of ruthenium oxide species, the reduction temperature of nickel oxide shifted to the lower temperature at 546 °C for RuO$_x$/Ni$_3$Al$_1$-LDOs compared with Ni$_3$Al$_1$- LDOs. The low reduction temperature may be caused by the interaction between ruthenium oxide species and nickel oxides [37].

Further comparing the reduction behavior of $RuO_x/Ni_3Al_1$-LDOs with $RuO_x/Al_2O_3$, the reduction temperature of ruthenium oxide species on the former was lower than that of the latter, indicating that the stronger ability to activate $H_2$ for $RuO_x/Ni_3Al_1$-LDOs. Based on the results of XRD and TEM, the low reduction temperature of ruthenium oxide species on $RuO_x/Ni_3Al_1$-LDOs could be attributed to the ultrafine ruthenium oxide species derived from the enhanced metal-support interaction.

The surface acidity of the prepared catalysts was determined by $NH_3$-TPD experiments (Fig. 4b). The $NH_3$ desorption peaks presented on the $NH_3$-TPD patterns of the prepared catalysts can be associated with three kinds of acid sites with different acidic strength, including weak acid sites ($W_A$, < 200 °C), moderate acid sites ($MS_A$, 200 °C-500 °C), strong acid sites ($S_A$, > 500 °C) [38]. All the prepared catalysts appeared the $W_A$ sites at about 90 °C. $Ru/Ni_3Al_1$-400 appeared the obvious $MS_A$ sites at about 368 °C, and only a small $NH_3$ desorption peak was observed at 232 °C for $Ru/Al_2O_3$-400. The surface acidity data determined by $NH_3$-TPD was summarized in Table S2. $Ru/Ni_3Al_1$-400 showed more acidity than $Ru/Al_2O_3$-400. In short, the flower-shaped morphology and highly dispersed Ru nanoparticles on $Ru/Ni_3Al_1$-400 could provide more accessible acid sites, which were beneficial for the hydrogenation reaction via accelerating the adsorption of VL and the breaking of C-O bond [36].

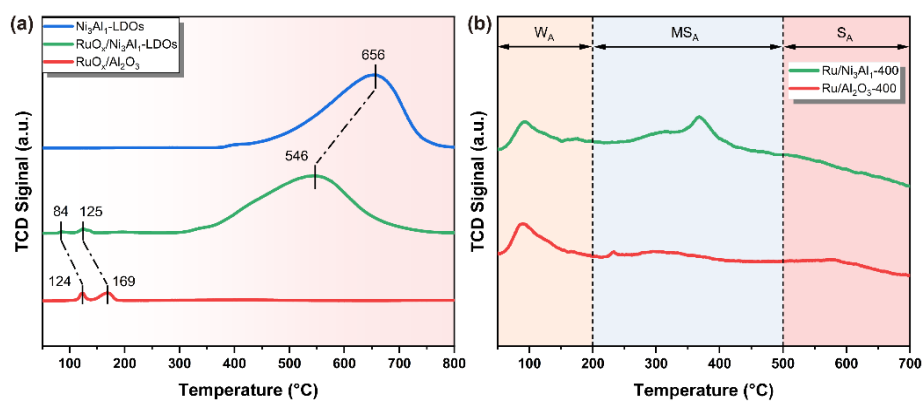

**Fig. 4.** $H_2$-TPR results of the different calcinated catalyst precursors (a). $NH_3$-TPD patterns of Ru/Ni$_3$Al$_1$-400 and Ru/Al$_2$O$_3$-400 (b).

## 3.2. Catalytic performance for VL HDO.

The HDO reaction of VL was used to evaluate the catalytic activity of the prepared catalysts. As shown in Fig. 5b1, the Ni$_3$Al$_1$-400 catalyst showed extremely poor catalytic activity with 2.9% yield of MMP at 73.8% conversion of VL. After the introduction of Ru species, Ru/Ni$_3$Al$_1$-400 displayed excellent catalytic activity, achieving complete conversion of VL and 94.2% yield of MMP, indicating that metal Ru played the role of active sites in the formation of MMP product. For comparison, Ru/Al$_2$O$_3$-400 only provided 51.8% yield of MMP under the same reaction conditions. Moreover, the turnover frequency (TOF) of Ru/Ni$_3$Al$_1$-400 and Ru/Al$_2$O$_3$-400 was calculated based on the Ru species (Table S3). Ru/Ni$_3$Al$_1$-400 exhibited a higher TOF value (57.8 h$^{-1}$) than Ru/Al$_2$O$_3$-400 (25.2 h$^{-1}$), which further indicated the superior catalytic activity of the former for VL HDO. Table S4 confirmed that Ru/Ni$_3$Al$_1$-400 exhibited comparable yield of MMP with the reported catalysts in previous references for the HDO of VL. The superior catalytic activity of Ru/Ni$_3$Al$_1$-400 might be due to the highly dispersed Ru species and abundant acid sites on LDHs-based support. The catalytic activity of RuO$_x$/Ni$_3$Al$_1$ sample reduced in H$_2$ flow at different temperature (500 °C, 600 °C) was also tested. The XRD pattern of Ru/Ni$_3$Al$_1$-500 and Ru/Ni$_3$Al$_1$-600 was shown in Fig. S3. The catalytic performance of Ru/Ni$_3$Al$_1$-500 and Ru/Ni$_3$Al$_1$-600 were similar to Ru/Ni$_3$Al$_1$-400. Therefore, Ru/Ni$_3$Al$_1$-400 was chosen as the study catalyst to explore the optimal reaction conditions for the HDO of VL over Ni$_3$Al$_1$-LDH-based catalyst.

The catalytic performance of Ru/Ni$_3$Al$_1$-400 at different temperatures was shown in Fig. 5b2. Even at low reaction temperature of 90 °C, Ru/Ni$_3$Al$_1$-400 achieved 96.4% conversion of VL. The main product was vanillin alcohol (VA) with 48.6% yield due to the relatively low temperature, and the yield of MMP was only 21.3%. VL could be completely converted while the reaction temperature was higher than 90 °C (100 °C-130 °C). The produced VA could be further converted into MMP. Ru/Ni$_3$Al$_1$-400 provided 99.9% conversion of VL and 94.2% yield of MMP at 130 °C. The effect of reaction time on the catalytic activity of Ru/Ni$_3$Al$_1$-400 was also investigated (Fig. 5b3). The reaction time of 2 h was not enough for the VL converted into MMP at 130 °C. From 4 h to 8 h, the primary product was MMP, and there was almost no VA produced. The effect of reaction pressure on catalytic activity was shown in Fig. 5b4, the yield of MMP increased with the increase of pressure, and only a little VA produced under the pressure from 0.1 to 0.5 MPa.

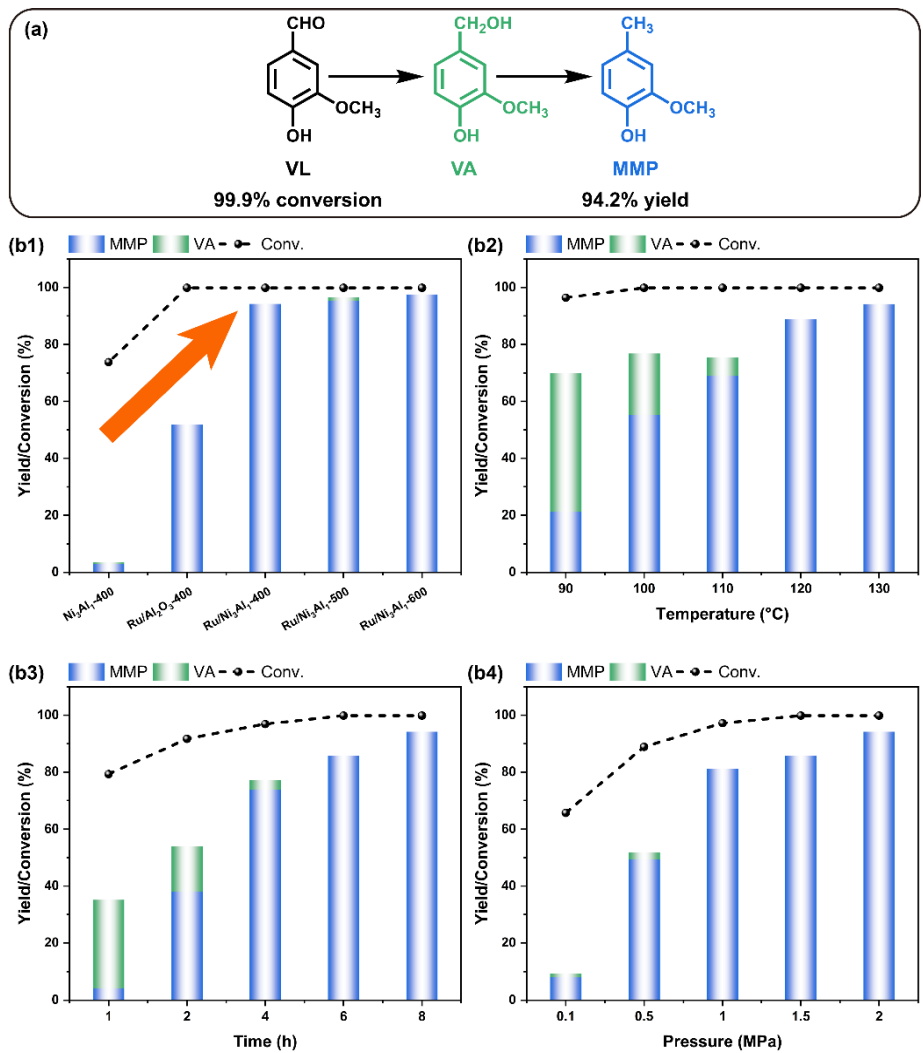

**Fig. 5.** Selective HDO of VL to MMP on Ru/Ni$_3$Al$_1$-400 (a). HDO performance of VL on different catalysts. Reaction conditions: 20 mg catalyst, 1 mmol VL, 130 °C, 2 MPa H$_2$, 8 h, 5 mL methanol (b1). Effect of reaction temperature (2 MPa H$_2$, 8 h) (b2), time (130 °C, 2 MPa H$_2$) (b3), and pressure (130 °C, 8 h) (b4) on the catalytic activity of Ru/Ni$_3$Al$_1$-400.

### 3.3. Proposed mechanism for HDO of VL.

On the basis of the results of catalyst characterization and catalytic tests, the possible mechanism of VL HDO over Ru/Ni$_3$Al$_1$-400 with methanol as solvent was proposed in Fig. 6. The highly dispersed Ru nanoparticles on the support acted as the role of active sites for the activation of $H_2$. The adsorbed $H_2$ on the surface of catalyst was dissociated into active hydrogen atoms. The excellent $Ni_3Al_1$-LDHs-basd support facilitated the adsorption of VL molecules which were prone to be attacked by the active hydrogen atoms. The aldehyde group on VL was hydrogenated into hydroxyl group with the formation of VA intermediate. The reaction solvent of methanol played a key role in the conversion of VA intermediate. As shown in Fig. S4, the solvent effect on the HDO of VL was investigated. The catalytic performance was poor with 78.5% conversion and 56.3% yield of VA when 1,4-dioxane as reaction solvent. Similarly, when tetrahydrofuran as solvent, the yield of MMP was only 67.9% with 96.3% conversion. Besides, 20.7% yield of VA was still observed in the reaction products. On the contrary, the catalytic activity of Ru/$Ni_3Al_1$-400 in ethanol and methanol were obviously higher than that of 1,4-dioxane and tetrahydrofuran. The results revealed that the alcoholic solvents were prone to combine with VA intermediate lead to the rapid conversion of VA. 2-methoxy-4-(methoxymethyl) phenol (DMP) produced from the *o*-methylation of VA with methanol was detected by GC-MS (Fig. S5) [39]. DMP was the main by-product during the HDO of VL with methanol as solvent. The solvent of methanol was more superior than ethanol for the production of MMP over Ru/$Ni_3Al_1$-400 due to the larger steric hindrance of the latter [40]. The newly formed DMP intermediate could be further attacked by the active hydrogen atoms, and then obtained the required product of MMP. Based on the reaction mechanism, there are three reaction pathways including hydrogenation, *o*-methylation, and hydrogenolysis for HDO of VL to MMP in methanol, which require the synergy of $H_2$ activation over metal sites and catalysis of acid sites on support. The catalysts characterization (XRD, TEM, XPS, and H$_2$-TPR) demonstrated the enhanced metal-support interaction over the LDHs-based catalyst, which improved the dispersion of Ru species on Ru/Ni$_3$Al$_1$-400, thus promoted the activation of H$_2$. Furthermore, the abundant acid sites existed in the LDHs-based support also facilitated the *o*-methylation and hydrogenolysis. Overview, the high yield of MMP for the HDO of VL over Ru/Ni$_3$Al$_1$-400 could be attributed to the synergistic catalysis of highly dispersed metal Ru centers and acid sites.

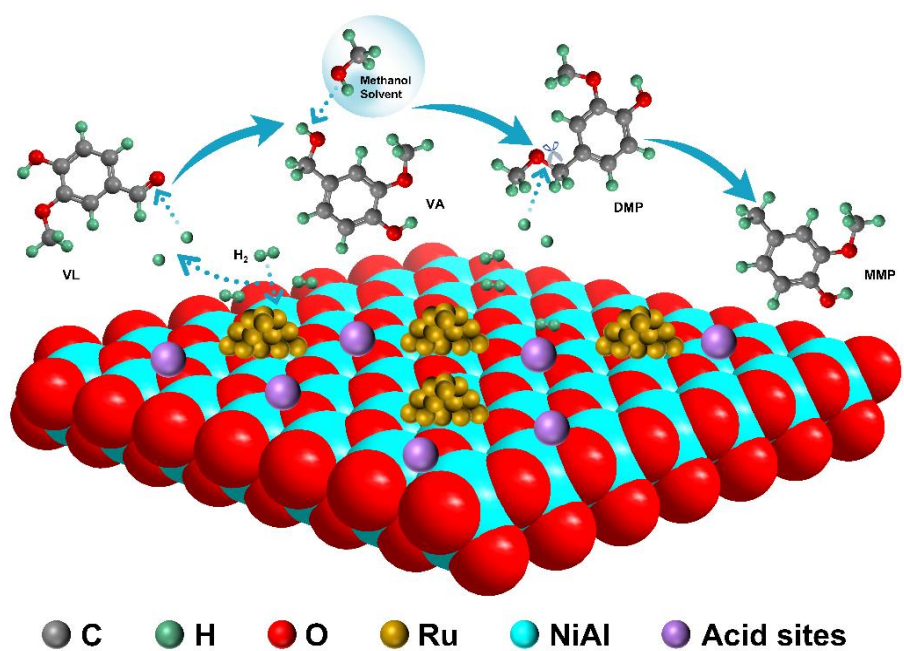

**Fig. 6.** Reaction mechanism of VL HDO over Ru/Ni$_3$Al$_1$-400 catalyst.

### 3.4. Catalytic stability test for HDO of VL.

The catalytic stability is an important index to evaluate the catalytic performance of the prepared catalysts. The recycle tests of Ru/Ni$_3$Al$_1$-400 for VL HDO were performed not only under the optimal reaction conditions (Fig. 7a) but also at lower VL conversion (~90%, Fig. S6). The reusability performance of Ru/Ni$_3$Al$_1$-400 confirmed that the catalyst could be used stably under the reaction conditions. The comparison of XRD and TEM of the fresh and used catalysts was used to further demonstrate the stability of Ru/Ni$_3$Al$_1$-400. As shown in Fig. 7b, the XRD of used Ru/Ni$_3$Al$_1$-400 showed no obvious difference from the fresh catalyst. Besides, no aggregation of Ru nanoparticles was observed on the used Ru/Ni$_3$Al$_1$-400 (Fig. 7c). The results of reusability revealed the good stability of Ru/Ni$_3$Al$_1$-400, which could be attributed to the enhanced interaction between Ru and support provided by the superiority of Ni$_3$Al$_1$-LDHs.

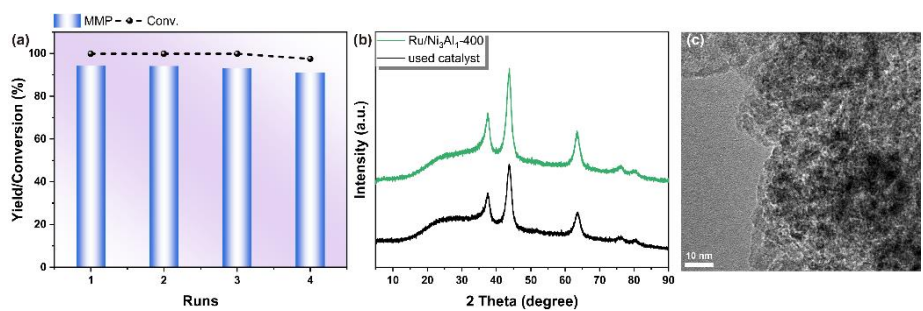

**Fig. 7.** Recycle tests of Ru/Ni$_3$Al$_1$-400 for VL HDO (a). Comparison of XRD patterns (b) and TEM images (c) of the fresh and used Ru/Ni$_3$Al$_1$-400 catalysts.

## 4. Conclusion

In summary, we constructed a Ni$_3$Al$_1$-LDHs-based support anchored highly dispersed Ru nanoparticles catalyst via a simple deposition-reduction method, which presented remarkable catalytic activity for the HDO of VL into MMP. The prepared Ru/Ni$_3$Al$_1$-400 catalyst achieved complete conversion of VL and 94.2% yield of MMP at 130 °C for 8 h with 2 MPa H$_2$ using methanol as solvent, which exhibited superior catalytic performance than the prepared Ru-free catalyst and Al$_2$O$_3$ supported Ru catalyst. A combined study indicates that the introduction of Ni$_3$Al$_1$-LDHs-based support enhances the interaction between metal and support and provides the highly dispersed Ru nanoparticles and abundant acid sites, which benefit the HDO of VL into MMP with the assistance of methanol solvent. This work provides a feasible strategy for the design of efficient catalyst with high selective toward biomass conversion.

**Acknowledgments**

This work was supported by the National Ten Thousand Talent Plan, National Natural Science Foundation of China (22078374, 21776324), Key Realm Research and Development Program of Guangdong Province (2020B0202080001), the Guangdong Basic and Applied Basic Research Foundation (2019B1515120058), Science and Technology Planning Project of Guangdong Province, China (2021B1212040008), Guangdong Laboratory for Lingnan Modern Agriculture Project (NT2021010), and the Scientific and Technological Planning Project of Guangzhou (202206010145).

964.